\documentclass[11pt]{article}

\usepackage{amsmath,amsfonts,amssymb,amsthm}
\usepackage{graphicx,color}
\usepackage{epsfig}
\usepackage{amssymb}
\usepackage{amsmath}
\usepackage[T1]{fontenc}
\usepackage{verbatim,url}
\usepackage[scaled]{beramono}
\usepackage{upquote}
\usepackage{algorithm}
\usepackage{algorithmic}
\usepackage{ifthen}

\usepackage{listings}

\definecolor{comments}{rgb}{0,.5,0}
\definecolor{backgnd}{rgb}{.95,.95,.95}
\definecolor{string}{rgb}{.2,.2,.2}
\title{Use of convexity in contour detection}
\author{Victor Churchill\\ \small Department of Mathematics, Dartmouth College}

\graphicspath{{./figs/}}

\begin{document}

\maketitle
\small
\subsection*{Abstract}
In this paper, we formulate a simple algorithm that detects contours around a region of interest in an image. After an initial smoothing, the method is based on viewing an image as a topographic surface and finding convex and/or concave regions using simple calculus-based testing. The algorithm can achieve multi-scale contour detection by altering the initial smoothing. We show that the method has promise by comparing results on several images with the watershed transform performed on the gradient images.

\normalsize
\section{Introduction}
In its first application, the watershed transform was performed on the gradient modulus of an image in order to detect contours around catchment basins, \cite{beucher1979use}. In this paper, a simple method for detecting similar contours is presented. In particular, contours at the boundary of convex and concave regions in the image are detected. This is achieved by locating zero crossings of Gaussian curvature in the image when viewed as a topographic surface. To identify depressions in the topography, convex regions of the surface are located. To identify protrusions, concave regions are located. In the proposed method these two region types can be viewed separately or together depending on the application.

The paper begins by considering convex regions in differentiable functions. In general, a twice differentiable function of $n$ variables $f$ is convex at a point if and only if Hessian matrix is positive semidefinite at that point, \cite{boyd2004convex}. Similarly $f$ is concave if and only if the Hessian matrix is negative semidefinite. This condition can be interpreted geometrically as the requirement that $f$ has positive (upward or downward, respectively) Gaussian curvature at each point $x$. If we were considering both convex and concave regions, this is where the surface has positive Gaussian curvature. For bivariate functions, these tests for convexity and concavity are performed by considering the behavior of the determinant of the Hessian matrix as well the second derivative with respect to the first variable. In order to formulate perform these convexity and concavity tests on a digital image, a pre-processing smoothing is performed so that the image better approximates a differentiable function. The size of the smoothing kernel determines the size of the features around which contours are formed.

The results show that this contour detection method provides several advantages over the watershed transform. The boundaries are always closed contours for segmentation, as it is impossible to move from from positive to negative Gaussian curvature without crossing zero. The method generally avoids oversegmentation that occurs with the watershed and any potential region-merging, user-defined markers \cite{meyer2012watershed}, or geodesic correction \cite{najman1996geodesic}. The other core advantage of this technique is algorithmic and code simplicity. Unlike the watershed transform, we don't need any notions of flooding, topographic distance \cite{meyer1994topographic}, or graph theory \cite{vincent1991watersheds}. The results are achieved using simple calculus-based testing. This enables a very fast computation. In our results, we compare with the watershed transform performed on the gradient modulus of the image of interest.

\section{Methods}
\subsection{Contour detection via convexity for functions}

Convexity of smooth bivariate functions can be determined by performing the second partial derivative test at every point in the domain. Images inherently have two dimensions, so it makes intuitive sense that the second partial derivative test for functions of two variables should have a connection with concavity in images.

Let $f(x,y)$ be a differentiable real-valued function of two variables whose second partial derivatives exist. The Hessian matrix of $f$ is
\begin{align}
H(x,y) &= \left[\begin{matrix}
f_{xx}(x,y) & f_{xy}(x,y)\\
f_{yx}(x,y) & f_{yy}(x,y)
\end{matrix}\right].
\end{align}
Define $D(x,y)$ as the determinant of $H(x,y)$
\begin{align}
D(x,y) := f_{xx}(x,y)f_{yy}(x,y)-(f_{xy}(x,y))^2.
\end{align}
The following conditions define convexity for bivariate functions.
\begin{enumerate}
\item If $D(x,y)>0$ and $f_{xx}(x,y)>0$ then $f$ is convex at $(x,y)$.
\item If $D(x,y)>0$ and $f_{xx}(x,y)<0$ then $f$ is concave at $(x,y)$.
\end{enumerate}
Note that these two conditions are equivalent to $H(x,y)$ being positive semidefinite and negative semidefinite, respectively. This can also be viewed as a condition on the Gaussian curvature, defined by
\begin{align}
K(x,y) = \frac{D(x,y)}{(1+(f_x(x,y))^2+(f_y(x,y))^2)^2}.
\end{align}
As the denominator is greater than zero, $\text{sgn}(D) = \text{sgn}(K)$. Hence if we wish to look at convex and concave regions as a single unit, we only need to find $D(x,y)>0$, which corresponds precisely to regions with positive Gaussian curvature. Note that Gaussian curvature is rotation-invariant, so it suffices to consider partial derivatives in the cardinal directions.

\subsubsection*{Example}
\begin{figure}[h!]
\centering
\includegraphics[width=.49\textwidth]{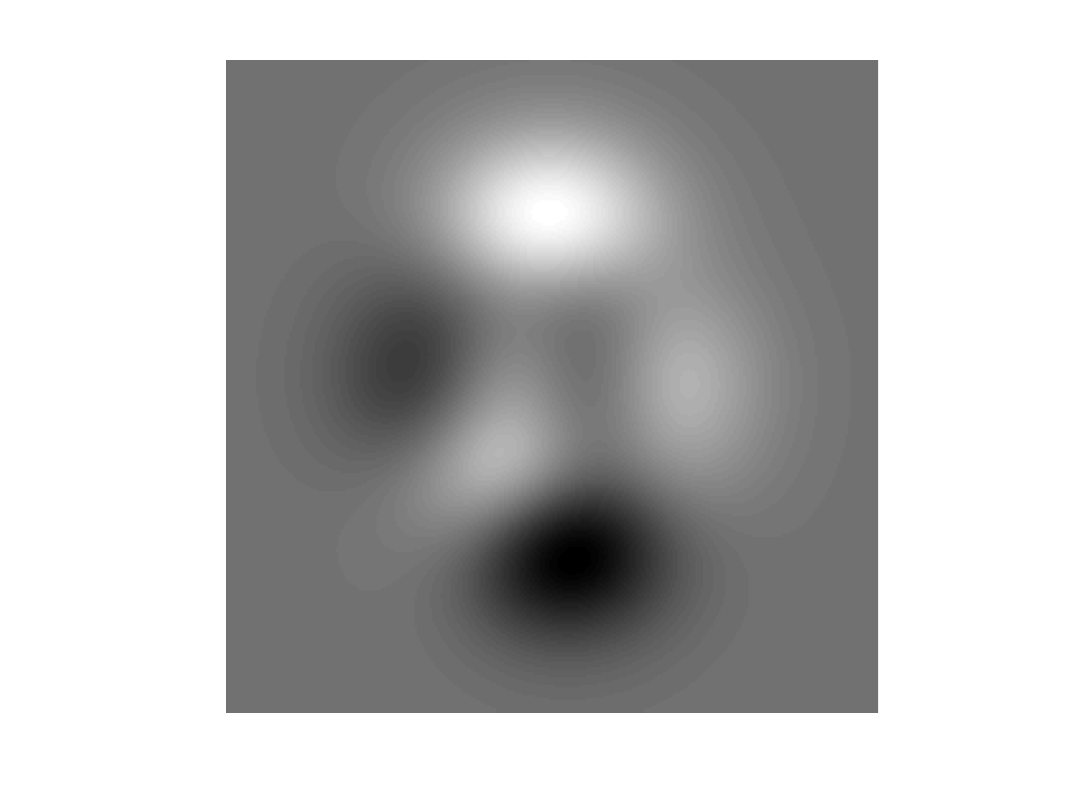}
\includegraphics[width=.49\textwidth]{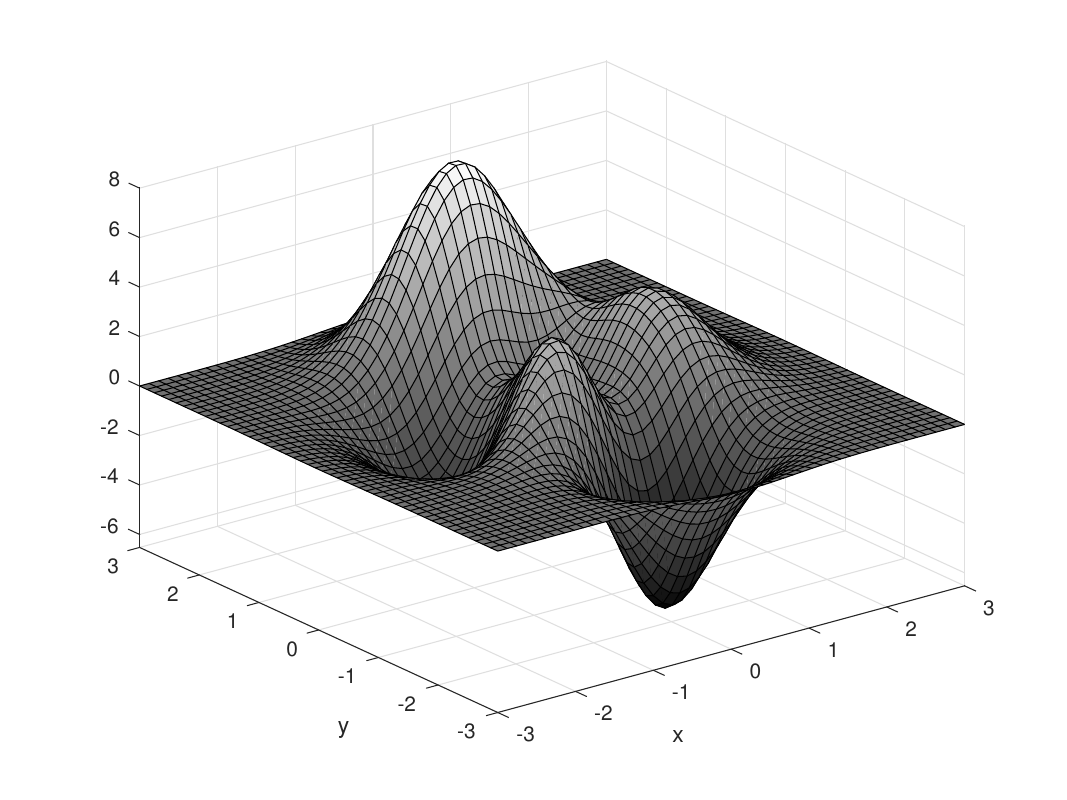}
\caption{Image and surface plot of $z(x,y)$.}
\label{fig:peaks}
\end{figure}

\begin{figure}[h!]
\centering
\includegraphics[width=.49\textwidth]{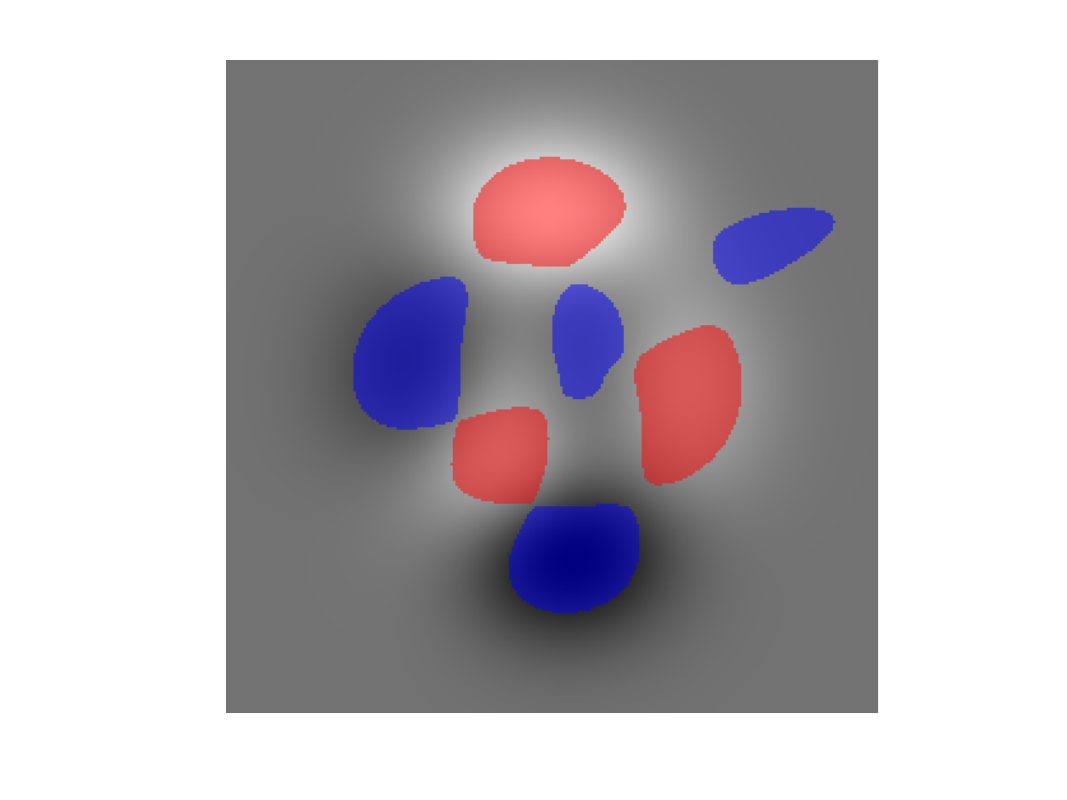}
\includegraphics[width=.49\textwidth]{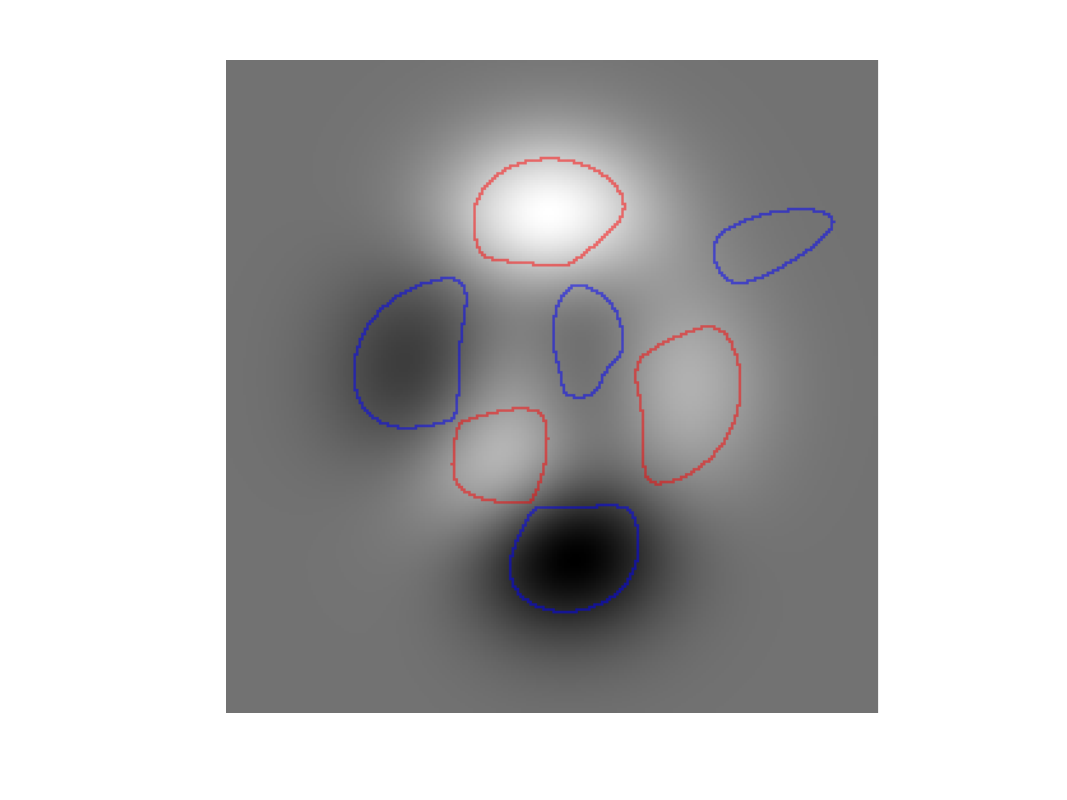}
\caption{Convex (blue) and concave (red) regions of $z(x,y)$ and their boundaries.}
\label{fig:peaksregions}
\end{figure}

As an example of finding convex and concave regions in a differentiable function of two variables whose second partial derivatives exist, we consider the function
\begin{align}
z(x,y) =  3(1-x)^2e^{-x^2 - (y+1)^2} - 2(x - 5x^3 - 5y^5)e^{-x^2-y^2} - \frac13 e^{-(x+1)^2 - y^2}.
\end{align}
The function and its topographic surface are shown in Figure \ref{fig:peaks}, and its convex and concave regions and their boundary contours are shown in Figure \ref{fig:peaksregions}. Note that the majority region where the function is neither convex nor concave corresponds to negative Gaussian curvature. That is, in these areas the principal curvatures are of differing signs.

\subsection{Contour detection via convexity for images}

Digital space is fundamentally different than function space. There are no infinitely small elements that are required for calculus. Hence an image needs to be pre-processed before the convexity test can be performed. Without pre-processing, the test typically generates very small scale closed contours which are not typically useful. Here, this is addressed by performing a smoothing whereby the image is convolved with a Gaussian kernel defined by
\begin{align}
k(x,y) = \frac{1}{2\pi\sigma^2}e^{-\frac{x^2+y^2}{2\sigma^2}}.
\end{align}
The parameter $\sigma$ determines the size of the smoothing kernel. As $\sigma$ increases, increasingly large convex and concave regions are typically flattened and hence not detected by the test. Hence, as shown later in the results, there is an inherent opportunity to detect contours at multiple scales in the same image by performing boundary detection on two differently smoothed versions of the same image. This can be seen in Figures \ref{fig:nuclei} and \ref{fig:multiscale}. The algorithm for contour detection is written out fully in Algorithm \ref{alg:contour}.

\begin{algorithm}
\caption{Contour detection via convexity}
\label{alg:contour}
\begin{algorithmic}[1]
\STATE Smooth the image $f$ using a Gaussian kernel of size $\sigma$ to obtain $f^\sigma$.
\STATE Compute $D(x,y)$ at each pixel in $f^\sigma$.
\STATE If $(p,q)$ is a pixel such that $D(p,q)>0$ and $f^\sigma_{xx}(p,q)>0$, label it with a $0$. Otherwise, $1$.
\STATE Consider the exterior boundary of the region found in the previous step.
\end{algorithmic}
\end{algorithm}

\section{Results}\label{sec:results}
\begin{figure}[t!]
\centering
\includegraphics[width=.49\textwidth]{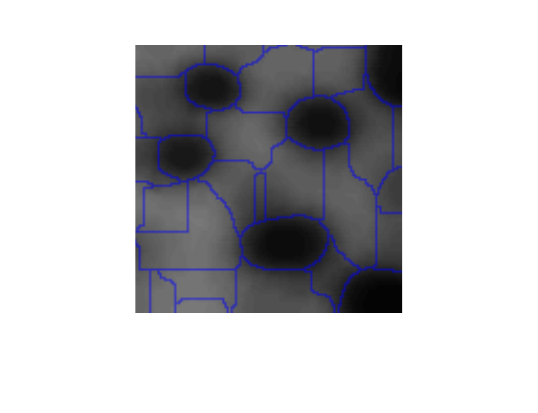}
\includegraphics[width=.49\textwidth]{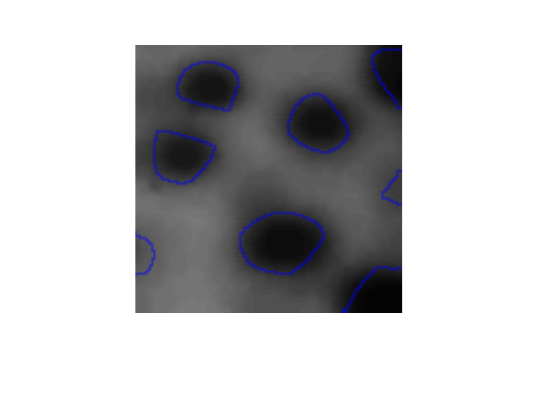}
\caption{Bubble detection in a radiographic plate \cite{watershedcmm} via watershed (left) and convexity (right).}
\label{fig:bubble}
\end{figure}

\begin{figure}[t!]
\centering
\includegraphics[width=.49\textwidth]{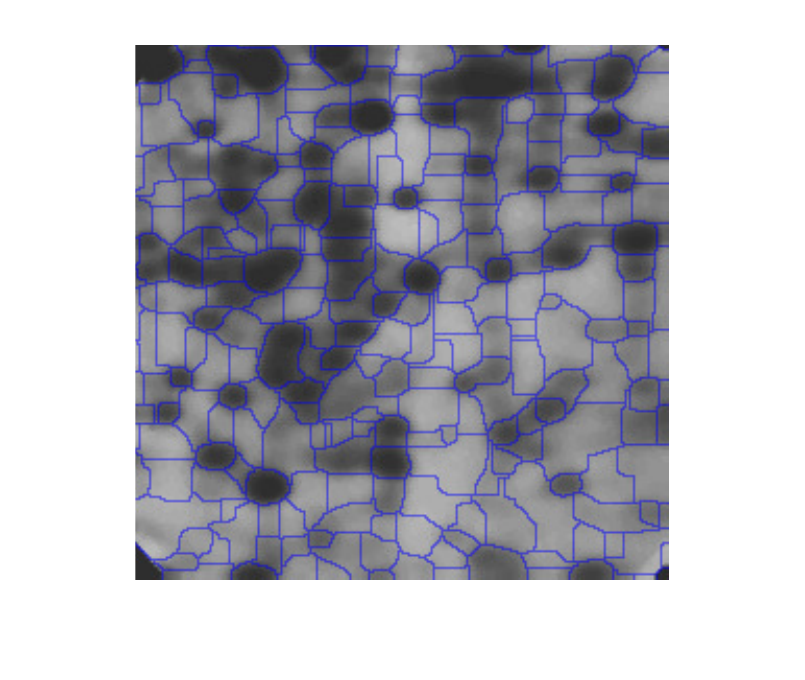}
\includegraphics[width=.49\textwidth]{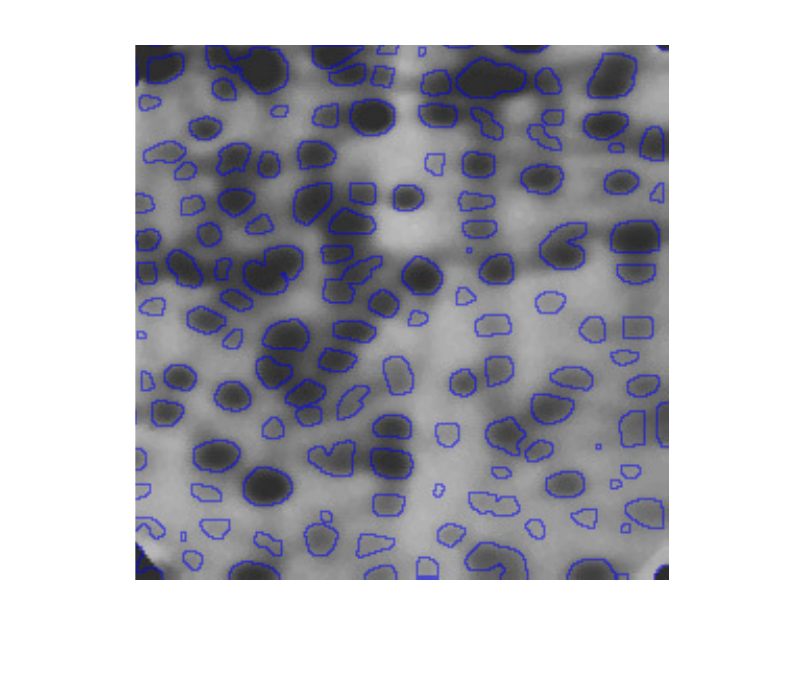}
\caption{Contours on an electrophoresis gel image \cite{watershedcmm} via watershed (left) and convexity (right).}
\label{fig:gel}
\end{figure}

\begin{figure}[t!]
\centering
\includegraphics[width=.49\textwidth]{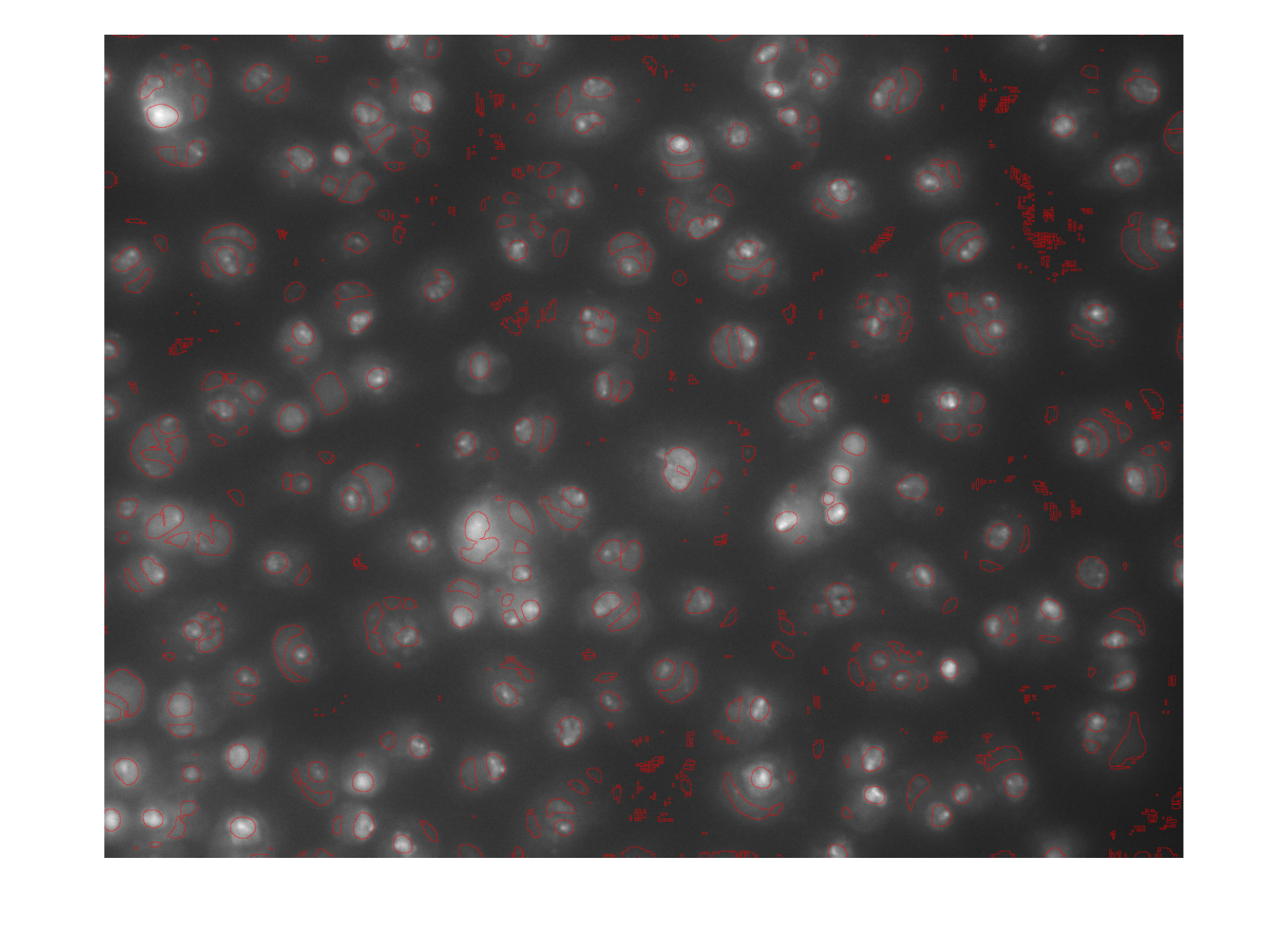}
\includegraphics[width=.49\textwidth]{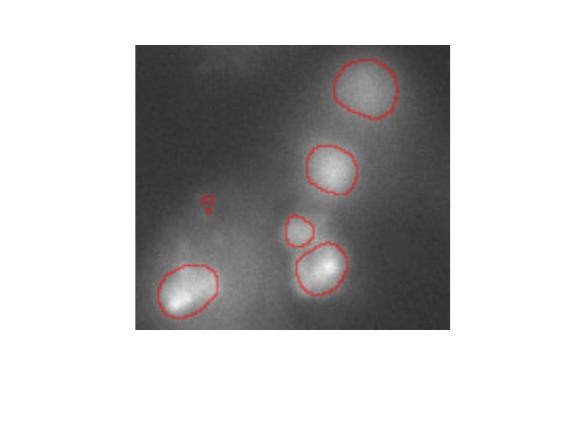}
\includegraphics[width=.49\textwidth]{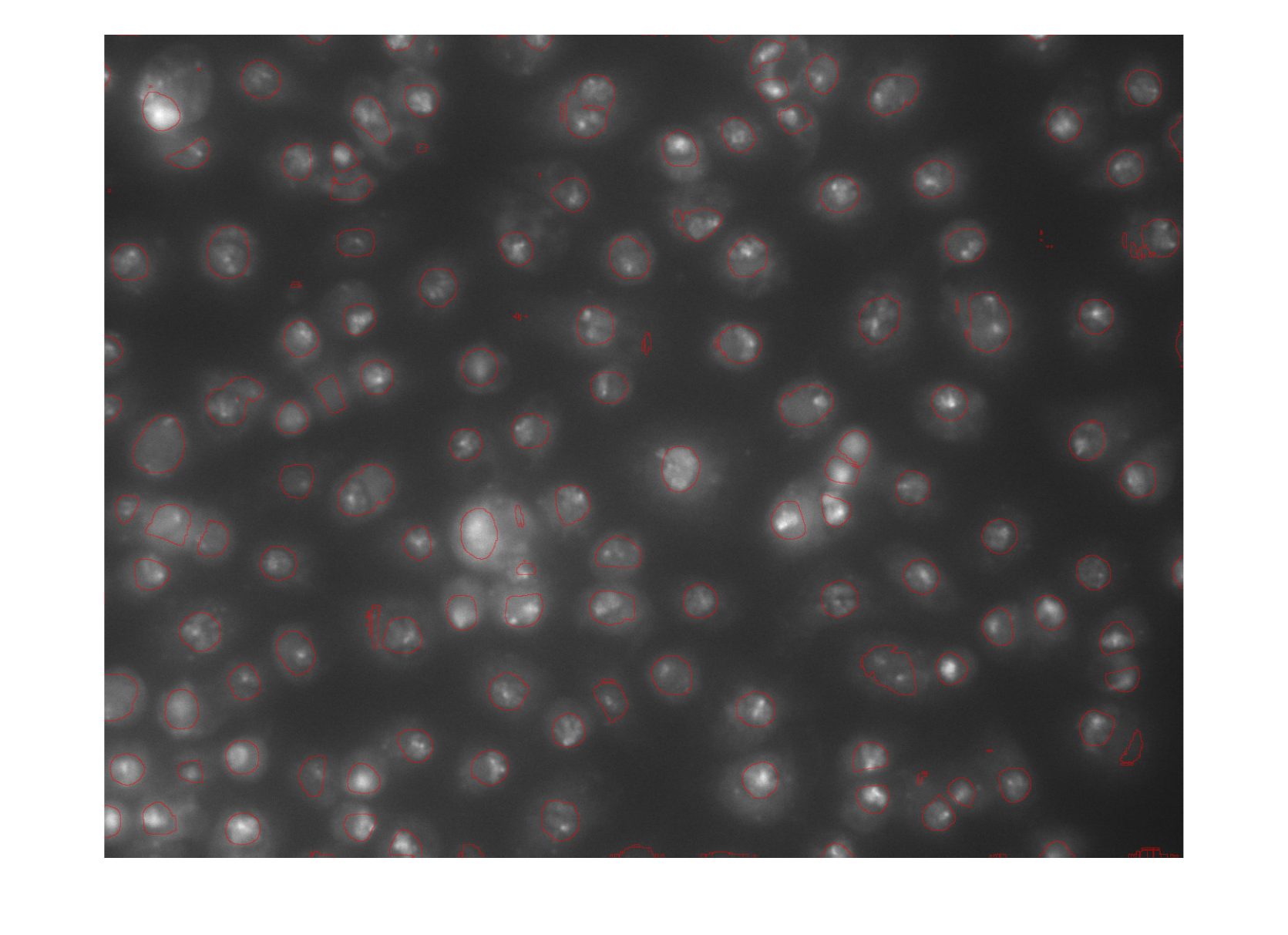}
\includegraphics[width=.49\textwidth]{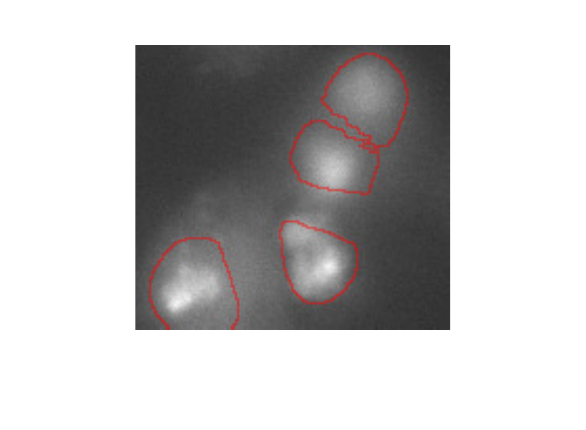}
\includegraphics[width=.49\textwidth]{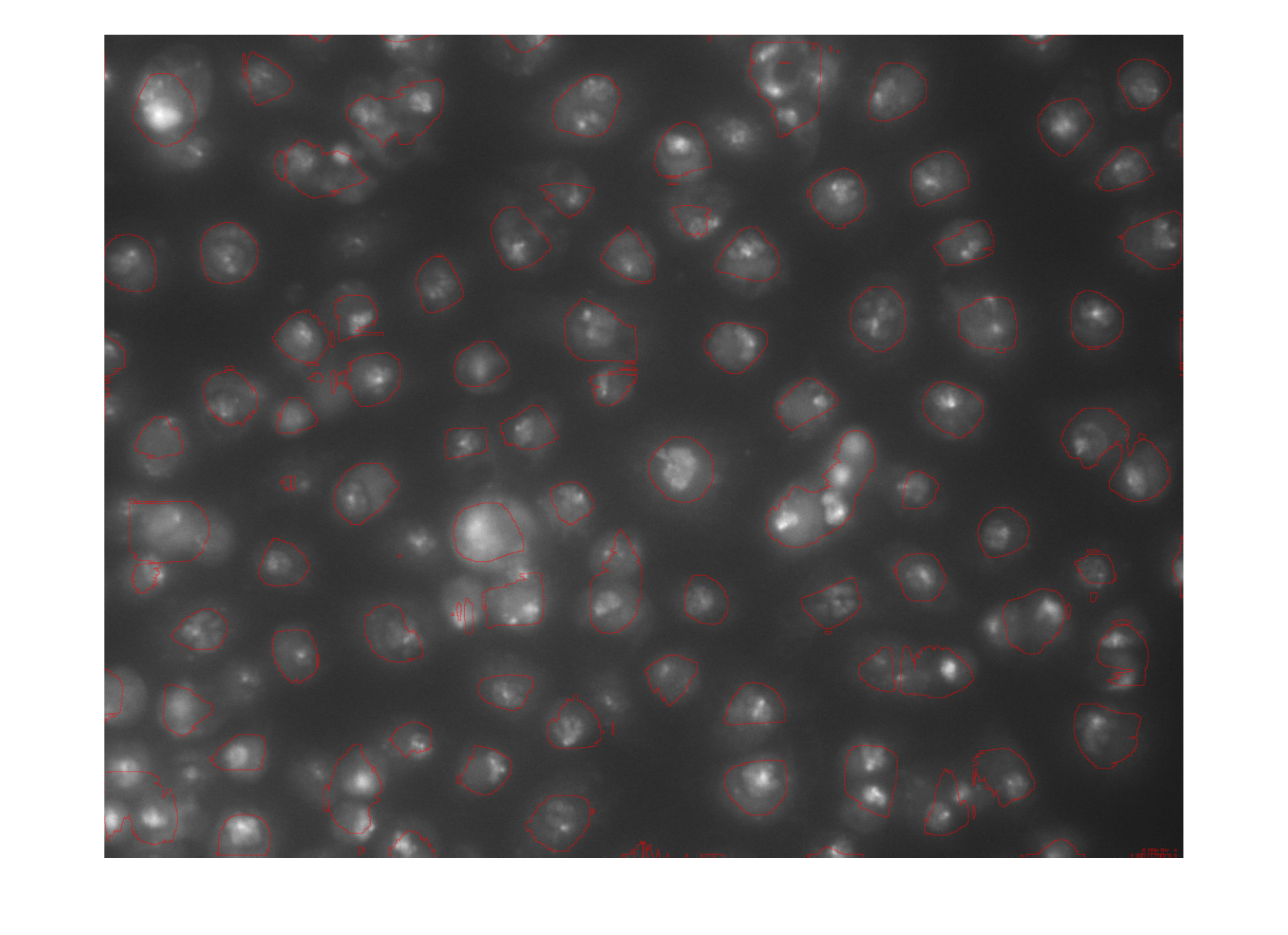}
\includegraphics[width=.49\textwidth]{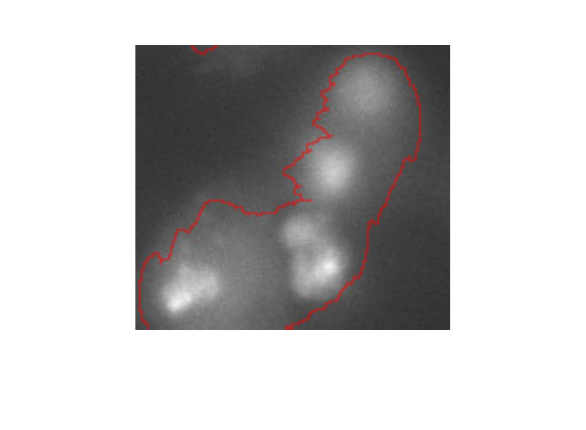}

\caption{Contour detection via convexity in an image of nuclei \cite{cellsegmentation} using three different-sized smoothing kernels. Right is a closeup of the full image on the left.}
\label{fig:nuclei}
\end{figure}

\begin{figure}[t!]
\centering
\includegraphics[width=.49\textwidth]{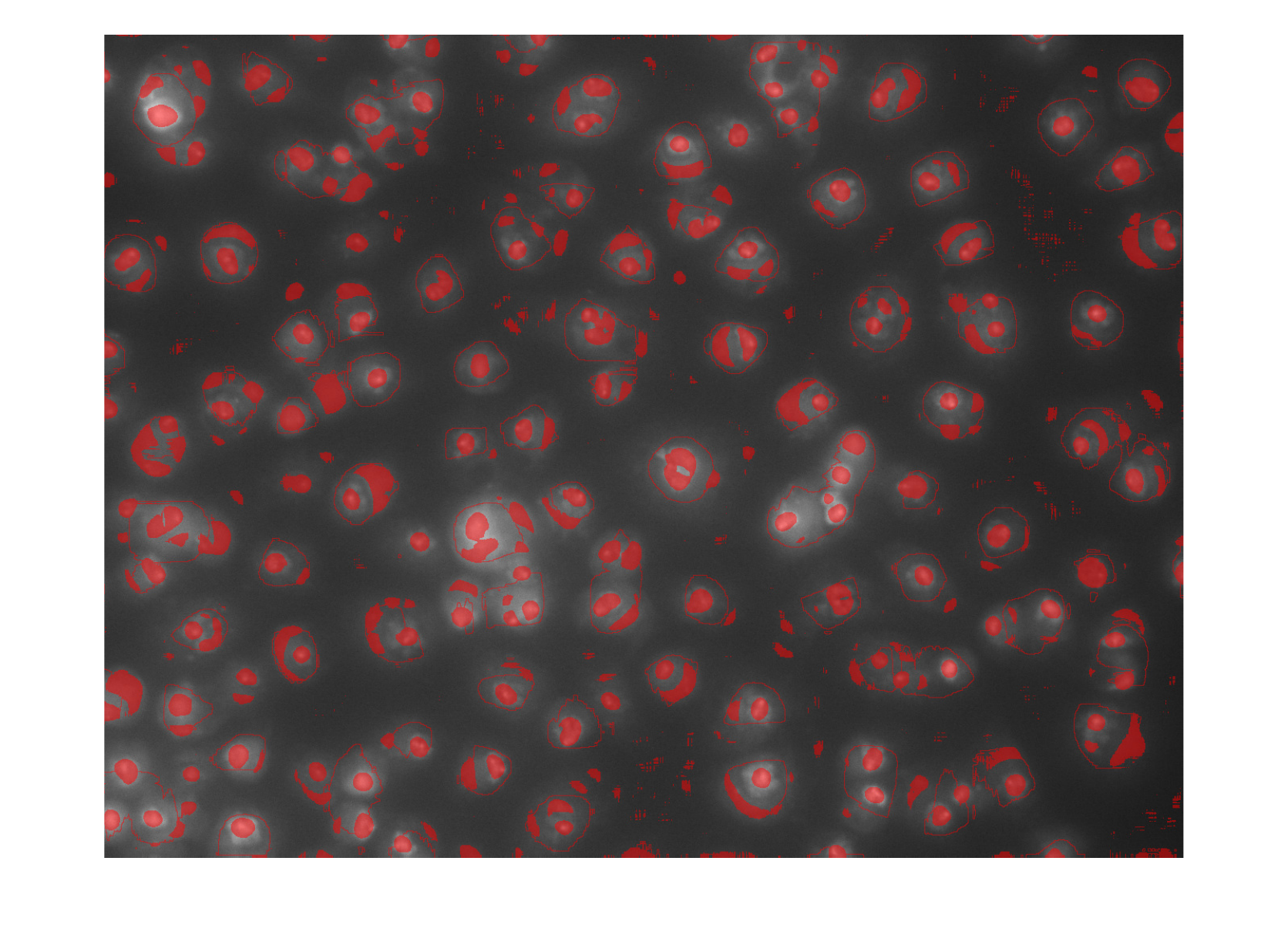}
\includegraphics[width=.49\textwidth]{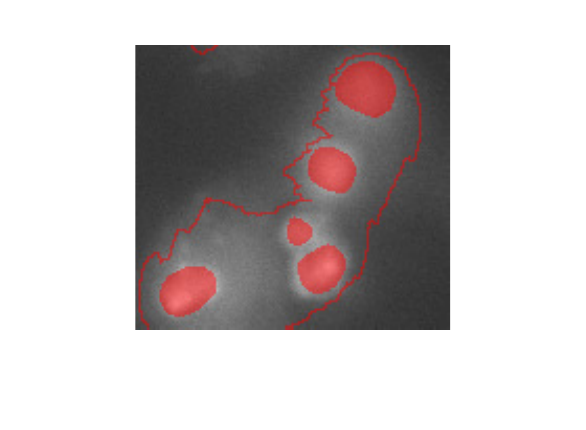}
\caption{Multi-scale detection via convexity in an image of nuclei \cite{cellsegmentation} using two different-sized smoothing kernels. Right is a closeup of the full image on the left.}
\label{fig:multiscale}
\end{figure}

\begin{figure}[t!]
\centering
\includegraphics[width=.49\textwidth]{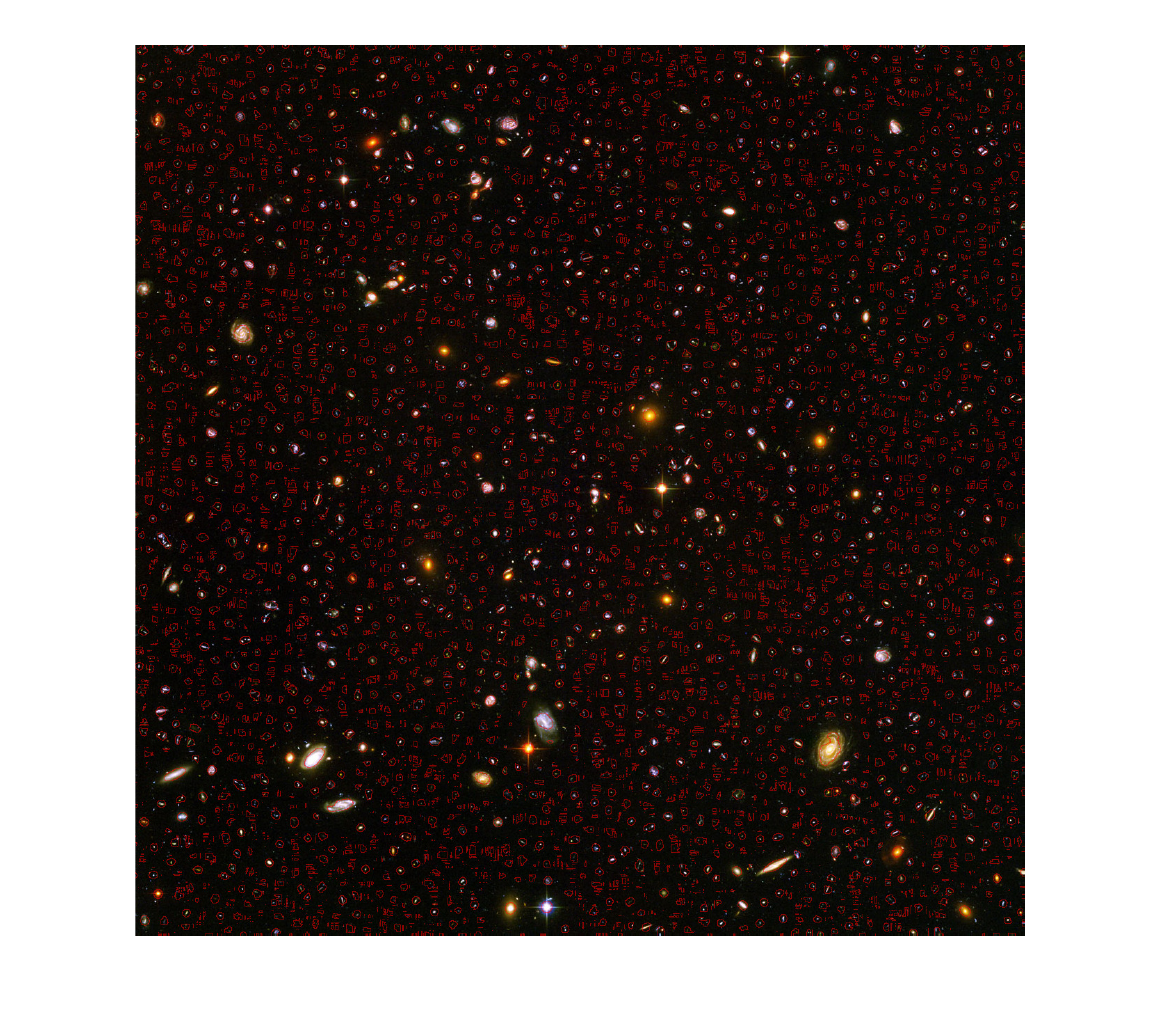}
\includegraphics[width=.49\textwidth]{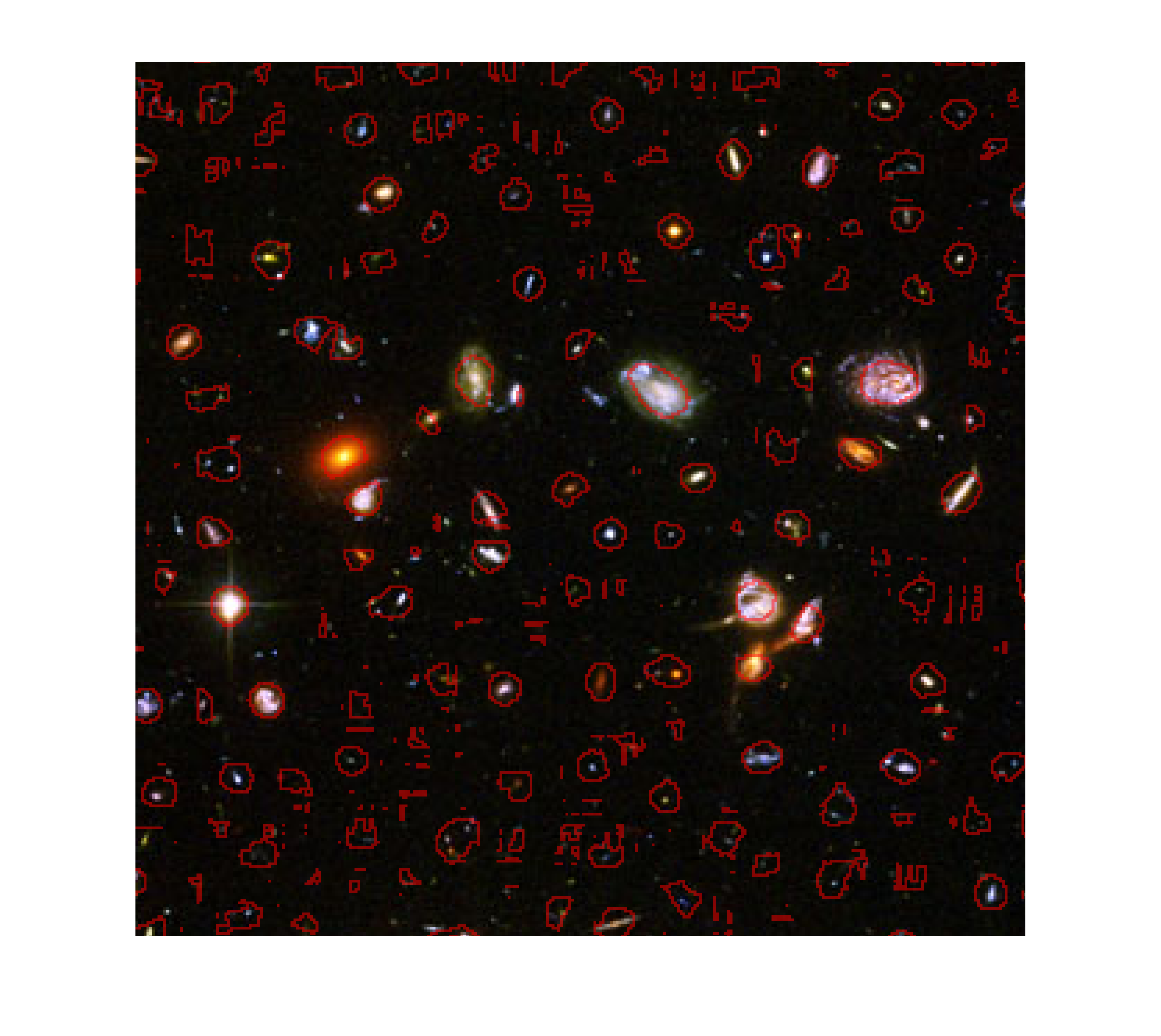}
\caption{Galaxy contour detection on an image from the Hubble Space Telescope, \cite{hubblespacetelescope_2014}. Right is a closeup of the full image on the left.}
\label{fig:galaxies}
\end{figure}

In this section, we compare the simple convexity-based contour detection algorithm with the watershed transform as formed in \cite{meyer1994topographic} applied to gradient images. In the original paper on watersheds, \cite{beucher1979use}, one application considered was bubble detection in a radiographic plate, \cite{beucher1992morphological}. This example is repeated in Figures \ref{fig:bubble}. We also look at detecting dark circular shapes in a gel electrophoresis image in Figure \ref{fig:gel}, where many more dark, and overlapping, spots are present. Figure \ref{fig:nuclei} shows the ability of the method to detect features of different scales using $\sigma=7.5,15,30$. Figure \ref{fig:multiscale} combines the $\sigma=7.5$ and $30$ results to depict multi-scale features of the original image. Small features are shown as filled in regions and large scales are shown as boundaries. Finally, Figure \ref{fig:galaxies} shows a large ($1200\times1200$ pixel) image of galaxies from which concave regions have been detected such that individual galaxies are identified by bounding contours.

\section{Conclusion and Future Work}\label{sec:conclusion}

This paper presented a contour detection algorithm that identifies convex and concave regions of an image. The results show several advantages of the proposed method for contour detection over the gradient watershed which we summarize below. First, the method typically does not oversegment if a reasonable smoothing parameter is chosen. Even if extraneous convex or concave regions are identified, they are always in the form of closed contours, and pruning based on size could be performed as post-processing. Foreground and background are much more clearly separated than by the gradient watershed. This method also has the advantage of providing multi-scale feature information in the image through the use of multiple smoothing kernels of different sizes. The simplicity of the method is superior both heuristically and in the code, an example of which is shown in the Appendix. This simplicity enables a very fast execution. A final advantage is how easily extended this method is to higher dimensions. As mentioned in the introduction, the convexity of an $n$-dimensional function is determined by the Hessian being positive semidefinite, or negative semidefinite for concavity. This simply-evaluated condition will allow contour and region detection on 3D and 4D datasets that have been appropriately smoothed as well. Finally, it is the author's hope that this technique will also aid in the problem of change detection, where a time-differenced image of the same scene can be smoothed and areas where change has occurred will be highlighted for further inspection.

\section*{Appendix - MATLAB Code Example}

\begin{lstlisting}
% load image of type double
im_original = imread('image');

% smooth image
sigma = 10;
im = imgaussfilt(im_original,sigma);

% compute Hessian determinant
[imx,imy] = imgradientxy(im);
[imxx,imxy] = imgradientxy(imx);
[~,imyy] = imgradientxy(imy);
D = imxx.*imyy - imxy.^2;

% find convex and concave regions
logical = D > 0;

% consider the exterior boundary
dilate = imdilate(logical,ones(3));
boundary = and(~logical,dilate);
\end{lstlisting}



\bibliographystyle{acm}
\bibliography{project}

\end{document}